\documentclass[11pt,nocontents]{iopart}

\usepackage{caption}
\usepackage[latin1]{inputenc}
\usepackage[T1]{fontenc}
\usepackage[english]{babel}
\usepackage{epsfig}
\usepackage{geometry}
\usepackage{bm}
\usepackage{graphicx}
\usepackage{ulem}
\usepackage{inputenc}
\usepackage{amssymb}
\makeindex 

\newenvironment{list2}{
  \begin{list}{$\bullet$}{%
      \setlength{\itemsep}{0in}
      \setlength{\parsep}{0in} \setlength{\parskip}{0in}
      \setlength{\topsep}{0in} \setlength{\partopsep}{0in}
      \setlength{\leftmargin}{0.2in}}}{\end{list}}

\begin{document}

\title{Galactic abundances as a relic neutrino detection scheme}

\author{Anna Sejersen Riis, Nikolaj Thomas Zinner, Steen Hannestad}
\address{Department of Physics and Astronomy, University of Aarhus \\
Ny Munkegade, DK-8000 Aarhus C, Denmark}

\ead{\mailto{asr@phys.au.dk},\mailto{zinner@phys.au.dk},\mailto{sth@phys.au.dk}}

\begin{abstract}
We propose to use the threshold-free process of neutrino capture on $\beta$-decaying nuclei (NCB) using all available candidate nuclei in the Milky Way as target material in order to detect the presence of the Cosmic neutrino background (C$\nu$B). By integrating over the lifetime of the galaxy one might be able to see the effect of NCB processes as a slightly eschewed abundance ratio of selected $\beta$-decaying nuclei. First, the candidates must be chosen so that both the mother and daughter nuclei have a lifetime comparable to that of the Milky Way or the signal could be easily washed out by additional decays. Secondly, relic neutrinos have so low energy that their de Broglie wavelengths are macroscopic and they may therefore scatter coherently on the electronic cloud of the candidate atoms. One must therefore compare the cross sections for the two processes (induced $\beta$-decay by neutrino capture, and coherent scattering of the neutrinos on atomic nuclei) before drawing any conclusions. Finally, the density of target nuclei in the galaxy must be calculated. We assume supernovae as the only production source and approximate the neutrino density as a homogenous background.
Here we perform the full calculation for $^{187}$Re and $^{138}$La and find that one needs abundance measurements with 24 digit precision in order to detect the effect of relic neutrinos. Or alternatively an enhancement of $\rho_{\nu}$ by a factor of $\sim10^{15}$ to produce an effect within the current abundance measurement precision.
\end{abstract}
\maketitle

\section{Introduction}
In analogy with the photons of the Cosmic Microwave Background (CMB), the neutrino component of the early Universe is expected to have decoupled from the hot primordial plasma once the rate of interactions with the rest of the particles dropped below the Hubble expansion rate \cite{Fuku:03}. The neutrinos travelling towards us from this last scattering surface are normally referred to as the Cosmic Neutrino Background (C$\nu$B) or relic neutrinos.

A detection of the C$\nu$B would be of great interest to cosmology as well as neutrino physics. Contrary to other neutrino sources outside the solar system, the C$\nu$B has the advantage of being a homogenous and isotropic background whose number density is only superseded by the CMB photon background. The theoretical expected number density is $n_{\nu + \overline{\nu}}$=112 cm$^{-3}$ for each neutrino mass state \cite{Fuku:03}. However, this advantage pales somewhat in comparison with the fact that the temperature of the background neutrinos is so low that their mean energy is many orders of magnitude below the threshold of any of the classical neutrino detection methods.

One can calculate the C$\nu$B temperature, seeing as it is closely related to the very accurately measured photon temperature, $T_{\gamma}=2.72548\pm0.00057$ K \cite{Fixsen:09}. Using entropy conservation, the photon temperature before reheating -- by electron-positron annihilation -- can be related to the temperature after reheating. The entropy is proportional to the number of degrees of freedom times the temperature cubed and therefore the final temperature is given by:
\begin{equation}
\centering
g_{*,f}T_f^3=g_{*,i}T_i^3.
\end{equation}
The number of degrees of freedom before reheating comes from both photons and electrons, so that $g_{*,i}=1+7/4$. After reheating the only source of radiation is photons, i.e.~$g_{*,f}=1$. By assuming the photon temperature before reheating to be equal to the neutrino temperature one gets:
\begin{equation}
\centering
T_{\nu}=\left(\frac{4}{11}\right)^{1/3} T_{\gamma}.
\end{equation}
Consequently the neutrino temperature today corresponds to a neutrino energy of only 6.5$T_{\nu}^2/m_{\nu}$ or 3.15$T_{\nu}$ for non-relativistic and relativistic neutrinos respectively, and a temperature of $T_{\nu}=1.95$K translates to an energy of $T_{\nu}k_B\sim$ 1.6 $\cdot 10^{-4}$eV\cite{Cocco:07}. This energy is far below the threshold for both the radio-chemical and Cherenkov neutrino detectors that have previously been used in experiments with solar, atmospheric, reactor, accelerator, and supernova neutrinos. However, it has long been known that neutrino capture on $\beta$-decaying nuclei (NCB) does not have a threshold (if the normal $\beta$-decay process is already spontaneously occurring in nature). Due to energy conservation, the $\beta$-particle will receive all the energy of the decay plus one neutrino rest-mass\footnote{Under the reasonable assumption that the daughter nucleus is so heavy that its recoil energy is negligible.}. This results in a small peak in the $\beta$-spectrum one neutrino mass above the theoretical endpoint, which is basically a smoking gun signature for the relic neutrino background unless non-standard model physics is introduced.

The size of the peak is a measure of the number of capture processes, which depends of course on the cross section of the reaction, but also on the C$\nu$B number density. Recent calculations show that even if one can, in principle, resolve the neutrino mass in the $\beta$-spectrum\footnote{The calculation in question was performed for the KArlsruhe TRItium Neutrino (KATRIN) experiment, which is expected to be the first in a new generation of $\beta$-decay experiments able to determine the neutrino mass with sub-eV sensitivity \cite{KAT:04}.}, the background density will have to be enhanced by roughly a factor $10^9$ in order to produce a detectable CNB signal \cite{Kaboth:10}. Or alternatively, the number density of the source material should be enhanced by the same factor.

\section{A relic neutrino detection scheme}
\label{sec:schemr}

Seeing as a $\mathcal{O}(10^9)$ source-enhancement of any of the upcoming $\beta$-decay experiments is not a viable option \cite{KAT:04,Monfardi:2006, Nucci:2010, Monreal:09}, we suggest consider the advantages of the NCB process in a rather different detection scheme.

Our approach to the problem consists in letting the detector material for the relic neutrino density be basically all the matter in the galaxy. The idea, in short, is to look at the effect of neutrino capture on selected nuclei, hoping that the relative abundances of selected $\beta$-unstable nuclei -- for which both the mother and daughter nuclei have extremely long lifetime -- will be detectably changed by the presence of the neutrino background\footnote{This is somewhat reminiscent of the detection scheme presented for supernova neutrinos by Lazauskas {\it et al.} in \cite{Lazauskas:09}}. That is, the 'detection' would consist in demonstrating an anomalous mother- to daughter- nuclei ratio compared with the case without neutrino induced decays. Put differently, in light of the enormous difficulty of detecting the relic neutrino background, we imagine the largest possible detector and calculate whether or not this extreme approach will help us achieve the goal.

The following sections will deal with the details of the calculation.

Section \ref{sec:nu_ca} presents our two proposed candidate nuclei and a calculation of the cross sections for NCB on these nuclei can be found in subsection \ref{ssec:nu_N}. However, since relic neutrinos have very low momentum and hence a macroscopic de Broglie wavelength, coherent scattering on the atomic electron clouds should be taken into account. In subsection \ref{ssec:nu_e} we calculate the relevant cross section for this scattering and compare it with the NCB cross section in order to evaluate the relative importance of this effect.

Section \ref{sec:Gal_Ab} deals with the production sites and rates for the candidate nuclei must be considered. The process contains several important steps:
\vspace{0.1cm}
\begin{list2}
\item Firstly, which stars are able to produce our choice elements? For simplicity we have used supernovae as the only source. It is established that some heavy elements can also be produced via radiative pressure processes in e.g.~Ap stars \cite{Babcock:60}, but it is fair to say that this is a subdominant site for nucleosynthesis.

\item The next question must be: how much of a given element can be produced in a supernova? To answer this we used the nucleosythesis simulations of \cite{Rauscher:02} and produced a simple average abundance plus errorbars, assuming a supernova mass range of 10 to 25 $M_{\odot}$.

\item Once we know how much source material is produced per supernova, we need to know how many supernovae (or rather progenitor stars massive enough to produce a core-collapse supernova) have been produced in our galaxy during its lifetime? This question can be answered with a reliable function for the galactic Star Formation Rate (SFR). The production rate of supernovae can then be constructed in a way that includes the matter distribution of the galaxy in order to yield the total distribution of target material.

\item Finally the excess abundance of the daughter nucleus can be calculated by folding the target material and neutrino distributions with the cross section and performing an integration over time and volume. In addition we need to calculate unperturbed abundances. These consists of an integration of the supernova distribution (each producing a specified number of both mother and daughter nuclei) over volume and time.
\end{list2}
\vspace{0.1cm}

It should be noted that the C$\nu$B is not the only neutrino background in our galaxy. The physical background contains also a diffuse supernova neutrino background and the diffuse gamma ray burst neutrino background. However, these backgrounds are expected to be much more energetic than the relic neutrinos, but also much more diffuse. Therefore it is a reasonable first approximation to ignore the mixture of neutrino sources and consider the C$\nu$B to be the only relevant neutrino background.

\section{Nuclear candidates}
\label{sec:nu_ca}
For a nuclei to be a suitable candidate in our detection scheme, both the mother and daughter nucleus must be very stable, with a lifetime comparable to that of the galaxy. Otherwise the signature we are looking for will be diluted by standard $\beta$-decays. This requirement in itself presents a problem because the cross section is inversely proportional to the product of the Fermi integral and the half life, $f t_{1/2}$ -- see Eq.~(\ref{eq:cross-sec}). If we consider the best case scenario of an allowed decay, then $\log ft$ will typically be in the lower end of the $\sim 4 - 9$ -interval \cite{Singh:98}. Assuming the half life to be comparable to the life time of the Universe -- i.e.~$\mathcal{O}(10^{17})$ s -- would then imply $f$ to be somewhere in the range of $10^{-8}-10^{-13}$. However, seeing as $\log f$ is proportional to the Q-value of the $\beta$-decay this leads us to the conclusion that -- as a second requirement -- one must have nuclei with very small decay energies in order to achieve both a long half life and a minimal $\log f t_{1/2}$ value \cite{Dunlap:04,Evans:82}. As it turned out we were only able to meet the requirement on stability on the mother and daughter nuclei.

In fact we only identified two candidates with both a very long half-life and a stable daughter nucleus. These are the first and second order, unique forbidden decays of $^{187}$Re and $^{138}$La:
\begin{eqnarray}
\centering
 \nu_e+^{187}\mathrm{Re} &\rightarrow & ^{187}\mathrm{Os} + \mathrm{e}^- \nonumber \\
  \nu_e+^{138}\mathrm{La} &\rightarrow &^{138}\mathrm{Ce} + \mathrm{e}^-\nonumber
\end{eqnarray}
All additional details are listed in Tables~\ref{tab:relic} and \ref{tab:relicRe}.

\begin{center}
\begin{table}
  \begin{tabular}[htb!]{ |l | c | c | c | r | }
    \hline
    Decay & Decay type & Q-value & Abundance$_M$ & Abundance$_D$  \\
     \hline \hline
    $\nu + ^{187}$Re $\rightarrow$ $^{187}$Os + $e^-$& $\beta-$ & 2.467 keV & 62.6\% & 1.6\% \\
        \hline
    $\nu + ^{138}$La $\rightarrow$ $^{138}$Ce + $e^-$& $\beta-$ & 1044.0 keV & 0.0902\% & 0.25\% \\
    \hline
      \end{tabular}
  \caption{Candidate nuclei for relic neutrino induced beta decay. Data taken from \cite{Cocco:07,Audi:03,Singh:98,Dunlap:04,Tanimizu:00}.
  \label{tab:relic}}
\end{table}
\end{center}

\begin{center}
\begin{table}
  \begin{tabular}[htb!]{ |l | c | c | c | r | }
    \hline
    Deacy & log ft & $\tau_{1/2}$ & Degree of Forbiddenness  \\
     \hline \hline
      $\nu + ^{187}$Re $\rightarrow$ $^{187}$Os + $e^-$& 11.28 & $1.3727\cdot 10^{18}$s & first, unique \\
    \hline
     $\nu + ^{138}$La $\rightarrow$ $^{138}$Ce + $e^-$& 18.0 & $9.3977\cdot 10^{18}$s & second, unique \\
     \hline
     \end{tabular}
  \caption{Candidate nuclei for relic neutrino induced beta decay. Data taken from \cite{Cocco:07,Audi:03,Singh:98,Dunlap:04,Tanimizu:00}.
  \label{tab:relicRe}}
\end{table}
\end{center}

\subsection{Neutrino-nucleus cross section}
\label{ssec:nu_N}
For the calculation of the neutrino-nucleus cross section we have used the following expression for the cross section times the neutrino velocity \cite{Cocco:07}:
\begin{equation}
\centering
\sigma_{NCB}v_{\nu}=\frac{2\pi^2\ln2}{\mathcal{A}\cdot t_{1/2}}.
\end{equation}
Here $\mathcal{A}$ is a description of the ratio of the shape factors involved in the $\beta$-decay (denoted $\beta$) and the neutrino capture process (denoted $\nu$):
\begin{equation}
\centering
\mathcal{A}=\int_{m_e}^{W_0}\frac{C(E_e^{'}, p_{\nu}^{'})_{\beta} p_e^{'} E_e^{'} F(E_e^{'},Z)} {C(E_e,p_{\nu})_{\nu} p_e E_e F(E_e,Z)} E_{\nu}^{'} p_{\nu}^{'} \mathrm{d}E_e^{'},
\end{equation}
where $W_0=m_e+Q_{\beta}-m_{\nu}$ is the maximally available electron energy in the $\beta$-decay,  $F(E_e,Z)$ is the Fermi function and C is the nuclear shape factor.

In the case of an allowed decay the cross section reduces to the following simple expression:
\begin{equation}
\centering
\sigma_{NCB}v_{\nu}= p_e E_e F(E_e,Z) \frac{2\pi^2\ln2}{f  t_{1/2}}.
\label{eq:cross-sec}
\end{equation}
And $f t_{1/2}$ can be derived easily from any table of log ft values.

\begin{figure}
\centering
\includegraphics[scale=0.4]{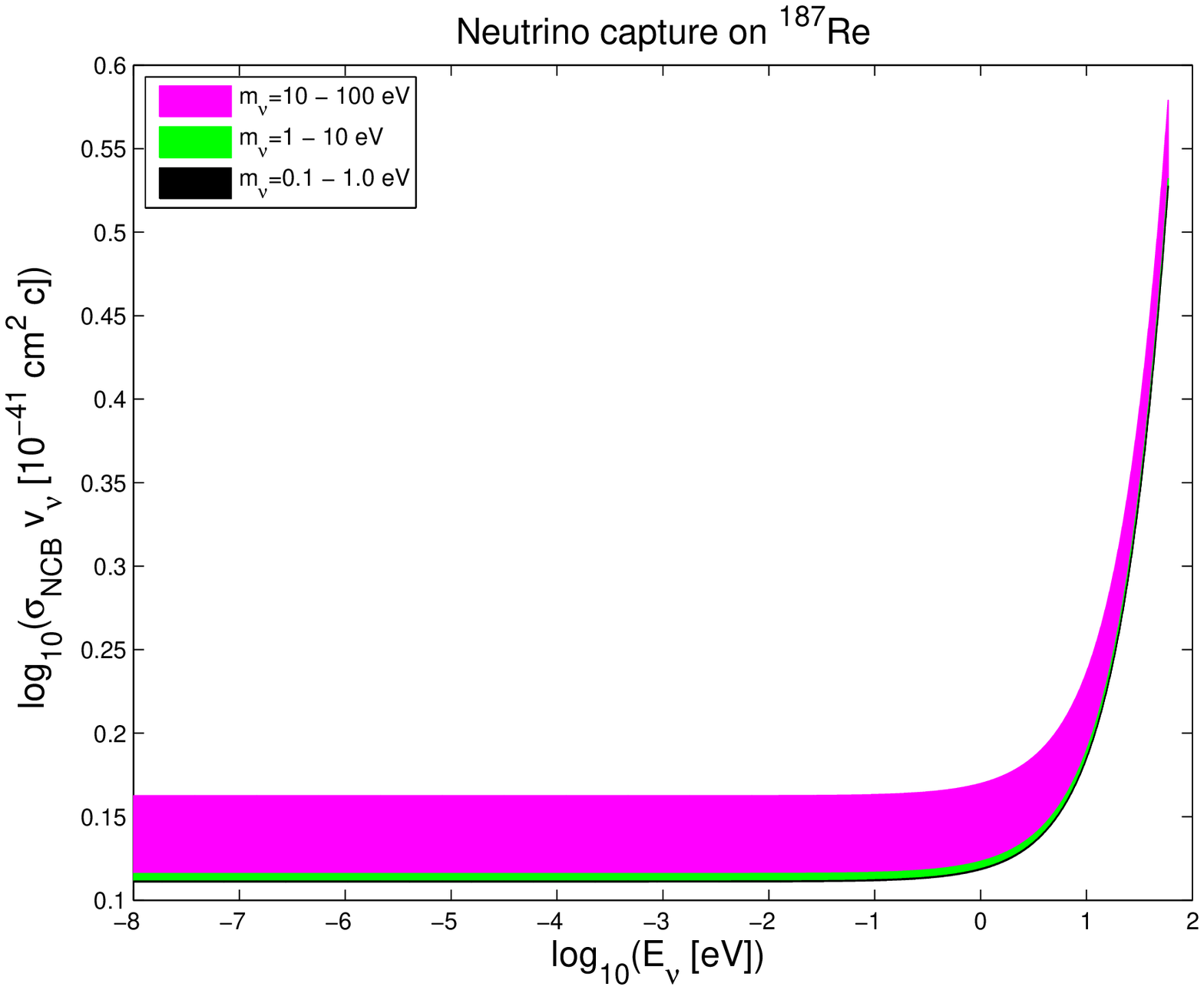}
\hfill
\includegraphics[scale=0.4]{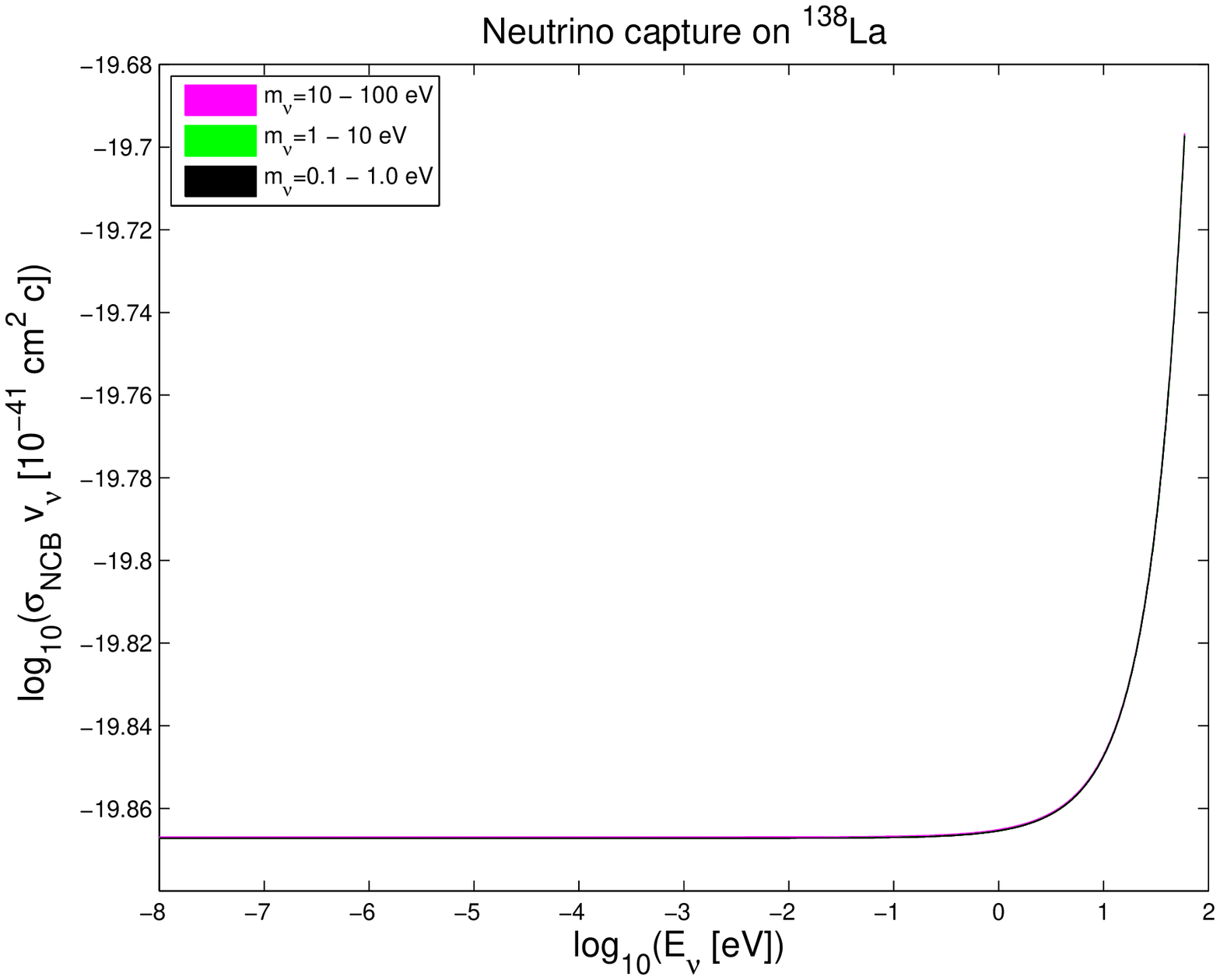}
\caption{The NCB cross section for Rhenium (left) and Lanthanum (right) as calculated for three different mass intervals. It is clear that only very large values of $m_{\nu}$ have any visible influence on the cross section of Re, while the La NCB cross section is practically not influenced by the neutrino mass at all. Note also that the results presented here are in good agreement with Figure 3 of \cite{Cocco:07}.  \label{fig:criss-cross}}
\end{figure}

However, for forbidden decays, the full machinery is needed. According to Behrens and B{\"u}ring, \cite{Behrens:82}, the shape factor can be simplified somewhat for unique K-forbidden processes:
\begin{equation}
\centering
C(E_e)=\frac{1}{3^2}R^4(^AF_{211}^{(0)})^2u_K,
\end{equation}
where the factors preceding $u_K$ will be divided out in the expression for $\mathcal{A}$. Meanwhile the relevant functions (for our purpose) are given by:
\begin{eqnarray}
\centering
u_1&=&p_{\nu}^2+\lambda_2p_e^2,  \\
u_2&=&p_{\nu}^4+\frac{10}{3}\lambda_2p_e^2p_{\nu}^2+\lambda_3p_e^4.
\label{eq:ups}
\end{eqnarray}
The details of the calculations of the coefficients can be found in \cite{Behrens:82}; for now, suffice it to say they depend on the {\it electron} energy and momentum and the nuclear charge and radius.

From a computational point of view, the main problem in the calculation of Eq.~(\ref{eq:ups}) is the presence of $\Gamma(z)$-functions in the $\lambda$-coefficients, which is to be calculated for very small energy input values (that is, very small $z$).

The cross sections for neutrino capture on Re and La are presented in Figure~\ref{fig:criss-cross} and agree nicely with the findings of Cocco {\it et al.} \cite{Cocco:07}.

\subsection{Neutrino-electron cross section}
\label{ssec:nu_e}

As mentioned in section \ref{sec:schemr} we are interested in the cross section for the scattering of neutrinos on atomic electrons:
\begin{equation}
\centering
\nu + e_{bound} \rightarrow  \nu + e_{continuum}
\label{eq:e_reaction}
\end{equation}

To calculate it we use the following expression, due to Bahcall \cite{Bahcall:64}:
\begin{equation}
\centering
\sigma=\frac{\sigma_0}{4}\int \mathrm{d}^3p |g(\mathbf{p})|^2\frac{(k^2-1)^2}{k^2}.
\end{equation}
Here $g(\mathbf{p})$ is the Fourier transform of the bound electron wavefunction and $k$ is the four-vector $(\epsilon_e + \epsilon_{\nu},\mathbf{p_e}+\mathbf{p_{\nu}}$), with $(\epsilon_{\nu},\mathbf{p_{\nu}})$ denoting neutrino energy and momentum and $(\epsilon_e,\mathbf{p_e})$ denoting electron energy and momentum. All these quantities are in units of $m_e$. Finally the front factor is defined as:
\begin{equation}
\centering
\sigma_0\equiv\frac{4}{\pi}\left(\frac{m_ec}{\hbar}\right)^{-4}\frac{G_F^2}{m_e^2c^4}=1.7\times 10^{-44}\mathrm{cm}^2.
\end{equation}

We assume that the scattering reaction of Eq.~(\ref{eq:e_reaction}) is only likely to happen with the weakest bound electrons. The ground state of atomic $^{187}$Re consists of 70 electrons in closed orbits and 5 electrons in the 5d (having a  binding energy of $-13.6$ eV). We therefore consider the average cross section for scattering with these five outer electrons. For $^{138}$La 56 electrons are in closed orbits and the last electron is again in the 5d orbital (with a binding energy of -7.3 eV) \cite{Lean:81}.

We now need an expression for the wave function for each of these target electrons. In principle a numeric expression can be calculated with e.g.~the electronic structure modeling programme GAUSSIAN \cite{gaussian}, enabling the user to choose a number of specifications, ranging from the class of the basis functions (used when expanding the wave function in polynomials) to the stability requirements of say, the radius or energy levels in question. However, for easy reference, and because we are more interested in the competition between the cross sections and less in the precise properties of the atom in question, we have used the Roothan-Hartee-Fock wave functions of \cite{Lean:81} to calculate the desired Fourier transforms.

In the Hartree-Fock approach to atomic structure a one-electron wave function or spin-orbital, $\phi$, can be constructed as an appropriately weighted sum of Slater functions\footnote{The total atomic wavefunction is given by the Slater determinant, which can be built from a combination of all the electronic wave functions: $\Phi=\mathcal{A}(\phi_1^{(1)}, ... , \phi_n^{(n)})$, where $\mathcal{A}$ denotes asymmetrization}:

\begin{equation}
\centering
\chi=[(2 n)!]^{-1/2}(2 \zeta)^{n+\frac{1}{2}} \hspace{2mm}r^{n-1}\exp(-\zeta r)Y_{l m}(\vartheta, \varphi),
\end{equation}
containing an orbital exponent $\zeta$, quantum numbers $n$, $l$, $m$ and the speherical coordinates $r$, $\vartheta$ and $\varphi$.

\begin{figure}
\centering
\includegraphics[scale=0.7]{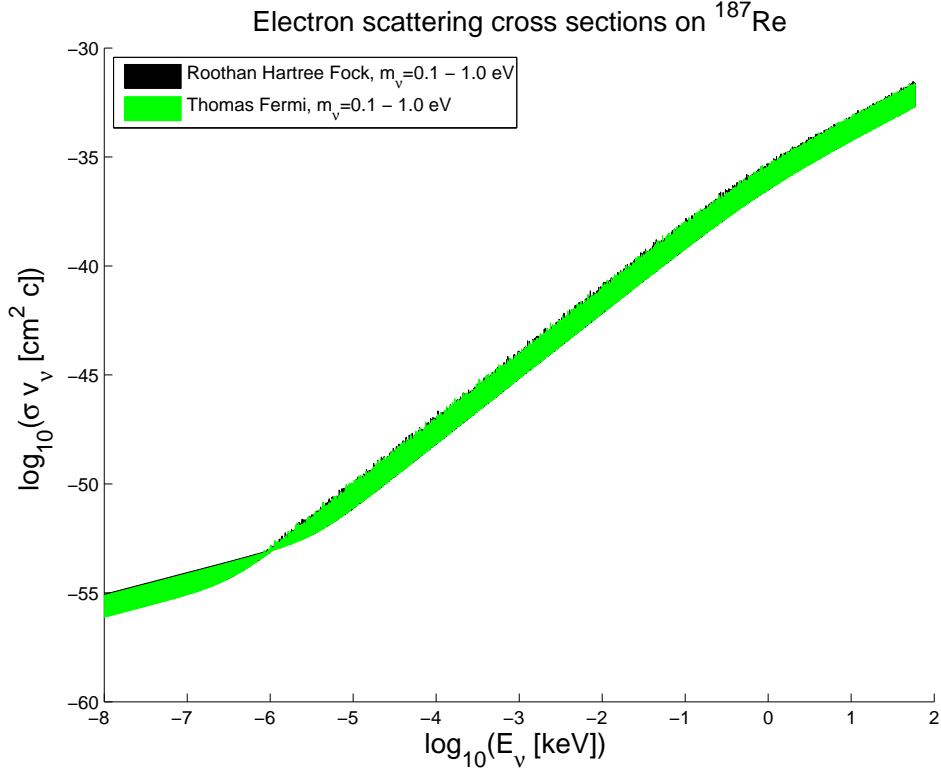}
\caption{A comparison of the cross section for Eq.~(\ref{eq:e_reaction}) as calculated with the Roothan Hartree Fock method of \cite{Lean:81} and a Thomas Fermi approximation for a range of neutrino masses. The two results are almost completely on top of each other, with the Thomas Fermi result in front. Although the Roothan Hartree Fock cross section is a bit higher than the Thomas Fermi cross section the results show a clear overall agreement. The average deviation between the two calculations lies between 17.5\% (for $m_{\nu}=0.1$ eV) and 16.6\% (for $m_{\nu}=1.0$ eV).  \label{fig:fermire}}
\end{figure}

As a consistency check we also performed the calculation of the cross section using a Thomas-Fermi approximation \cite{Joachain}. We found the two cross sections to agree reasonably well, given the nature of the approximations. Figure~\ref{fig:fermire} shows a comparison of the calculated cross sections.

\begin{figure}
\centering
\includegraphics[scale=0.4]{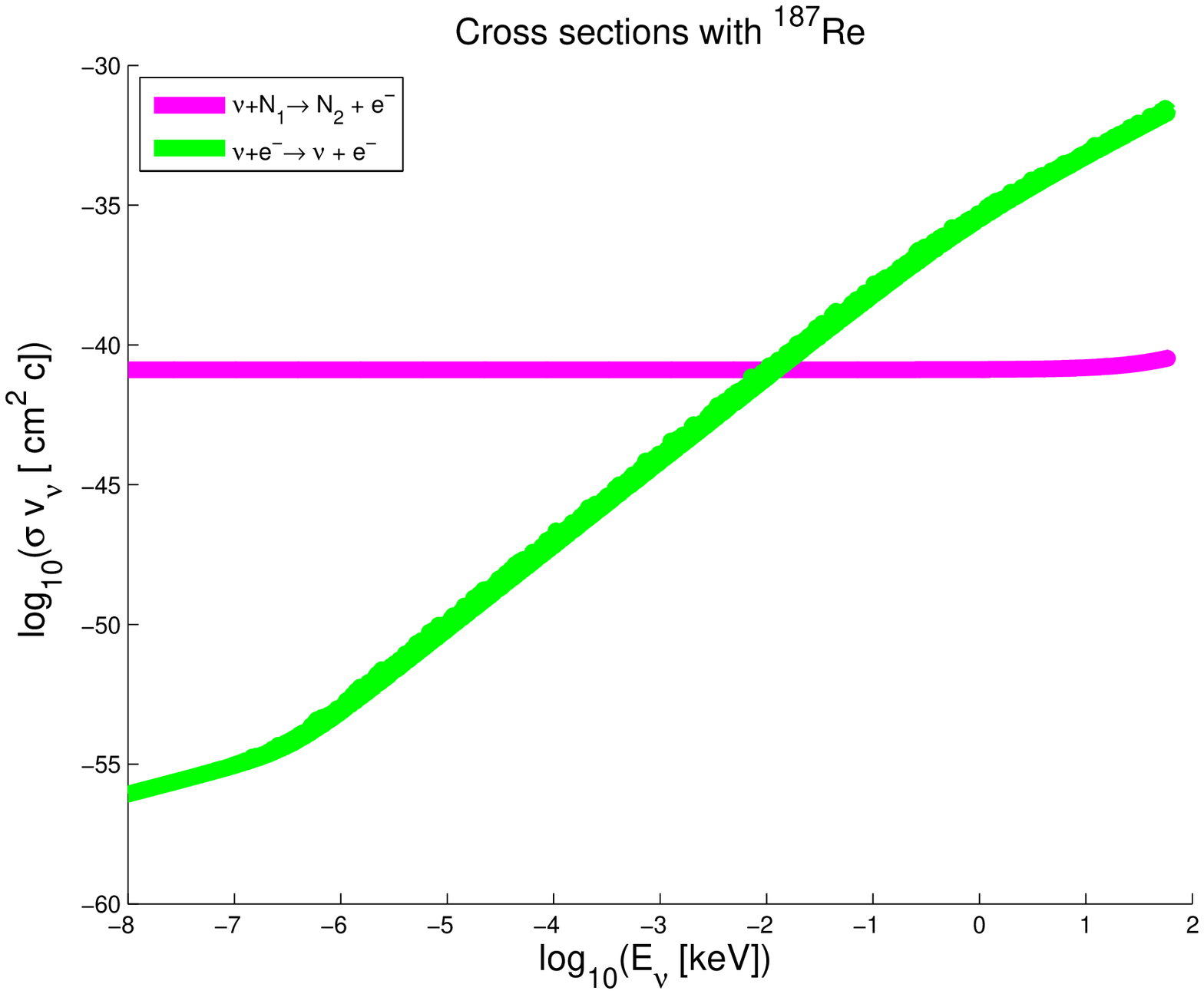}
\hfill
\includegraphics[scale=0.4]{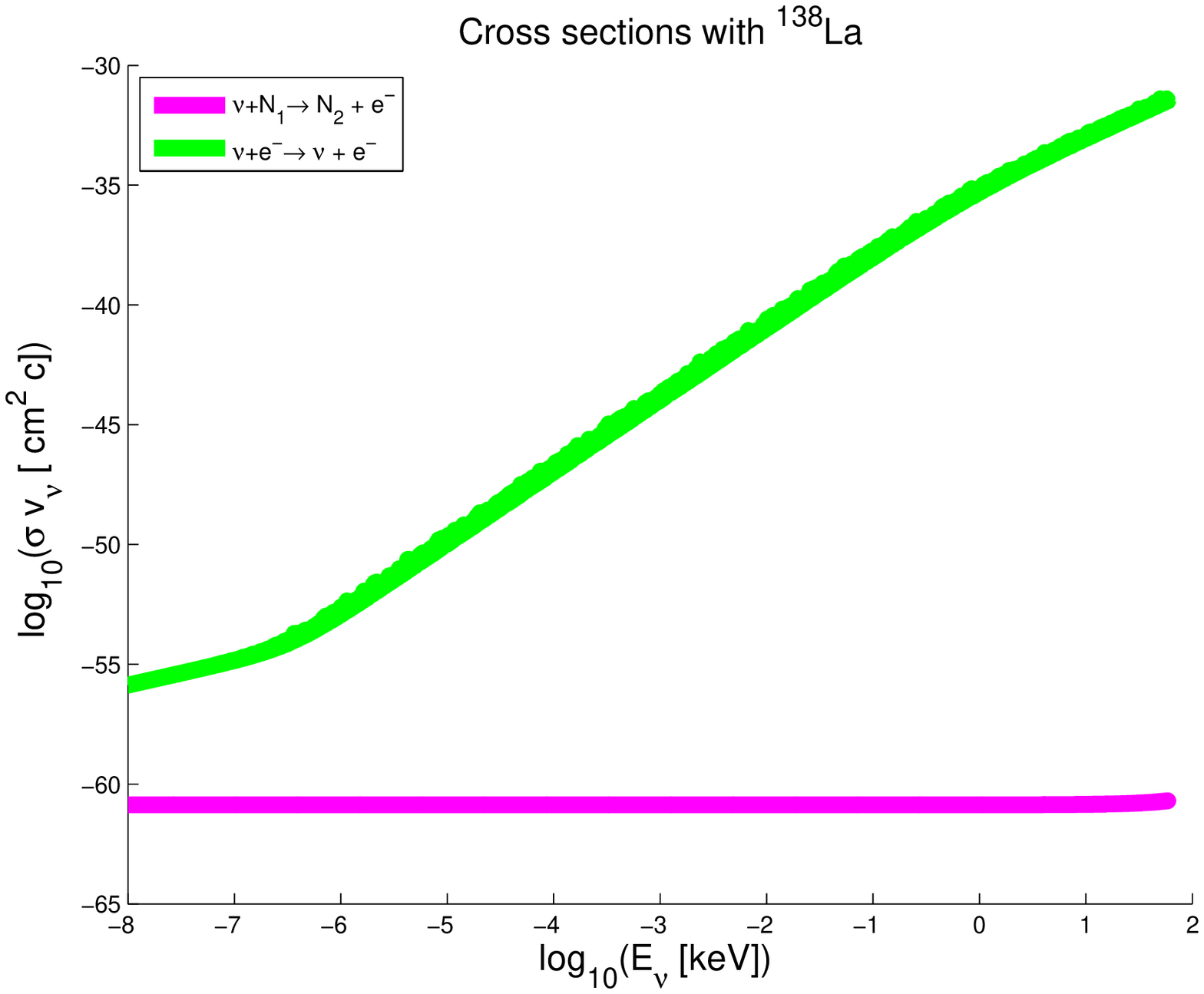}
\caption{A comparison of the NCB and coherent electron cross sections for Rhenium (left) and Lanthanum (right) for a neutrino mass of 0.1 eV. In the case of Rhenium the electron scattering becomes subdominant for neutrino energies below $\sim$ 10 eV which is well above any realistic relic neutrino energy. However, for Lanthanum the NCB cross section is consistently below the scattering cross section and we can conclude that neutrinos are far more likely to scatter on the electronic cloud than inducing any second order forbidden $\beta$-decays. \label{fig:criss-crosse}}
\end{figure}

Our results are presented in Figure~\ref{fig:criss-crosse}. For Rhenium the NCB cross section is fortunately by far the largest for energies less than $\sim$ 10 eV. So we can safely conclude that coherent scattering with atomic electrons does not disturb the neutrino capture process at the relevant relic neutrino energies. However the second order forbidden NCB process on Lanthanum has a cross section of order $10^{-60}$ which is far below the scattering cross section for the entire energy range, and we must conclude that this process is simply too weak to be useful in our detection scheme.

\section{Galactic Abundances}
\label{sec:Gal_Ab}
Having calculated the cross sections the NCB interaction rate we now evaluate :
\begin{equation}
\centering
\Gamma(t)= \int \sigma v_{\nu} n_{\beta}(z,R,t) \mathrm{d}z \mathrm{d}R,
\end{equation}
where $R$ and $z$ are the galactic radius and height coordinate, $t$ is the galactic time and $n_{\beta}(z,R,t)$ is the number density of the target nucleus. By assuming supernovae as the only production sites, the target nuclei density, $n_{\beta}$, can be expressed as the number of target atoms produced per supernova times the supernova rate:
\begin{equation}
\centering
n_{\beta}=N_{\mathrm{target, SN}}R_{\mathrm{SN}}
\end{equation}
If we consider the neutrino distribution as a homogenous background, our main objective -- the surplus abundance -- can now be calculated with a very simple expression:
\begin{equation}
\centering
A_{\mathrm{Re+, tot}}= n_{\nu} \int \Gamma \mathrm{d}t.
\label{eq:abund}
\end{equation}

The missing ingredient is thus $n_{\beta}$. In order to calculate it we need firstly an expression for the supernova rate. Fukugita and Kawasaki \cite{Fuku:02} provide the following expression:
\begin{equation}
 \centering
R_{\mathrm{SN}}= \psi(t)\frac{\int_{m_c}^{m_u} \phi(m)m^{-1} \mathrm{d}m }{\int_{0}^{m_u} \phi(m) \mathrm{d}m},
\label{eq:sn-rate}
\end{equation}
where $\psi(t)$ is the star formation rate (SFR) and $\phi(m)$ the initial mass function (IMF).

To evaluate $R_{\mathrm{SN}}$ we choose the traditional Salpeter IMF, which for stars heavier than the Sun equals $\phi(m)\sim m^{-1.35}$. As the upper and lower mass limits of Eq.~(\ref{eq:sn-rate}) we use $m_c=10M_{\odot}$ and $m_u=25M_{\odot}$, to ensure a mass range that includes core-collapse supernovae massive enough to actually produce our target materials, yet not so massive that the products of nucleosynthesis are swallowed by the ensuing collapse into a black hole \cite{Heger:03}. With these conservative choices we get:
\begin{equation}
\centering
R_{\mathrm{SN}}=0.00429 M_{\odot}^{-1}\psi(t).
\label{eq:rsn}
\end{equation}
Following \cite{Fuku:02} the time dependence of the star formation rate can be expressed as a simple exponential law:
\begin{equation}
\centering
\psi(t)=\psi(t_0)\exp[(t_0-t)/\tau],
\end{equation}
with $\tau=2.8$ Gyr for $z <1$ and $t_0$ taken to be the age of the Universe. The star formation rate is given by:
\begin{equation}
\centering
\log \psi(t_0) [M_{\odot} \mathrm{yr}^{-1} \mathrm{Mpc}^{-3}]=-2.09_{-0.13}^{+0.22},
\end{equation}

This expression has been derived in \cite{Fuku:02} using Eq.~(\ref{eq:sn-rate}) and the mean of three supernova rates (in units of $(10^{10}L_{B\odot}\cdot100 \mathrm{yr})^{-1}$) from three different populations of galaxies. To get the {\it local} star formation rate we make the following substitutions:
\begin{equation}
\centering
\psi(t_0)\rightarrow\frac{\psi(t_0)}{\mathcal{L}_B}\left(\frac{L_{\mathrm{MW},B}}{M_{\mathrm{MW},*}}\right)\rho_{\mathrm{MW}}.
\end{equation}
That is, firstly we divide $\psi(t_0)$ by the local B-band luminosity density of the universe, $\mathcal{L}_B=2.4\pm0.4\cdot10^8 h L_{\odot} \mathrm{Mpc}^{-3}$, (taken again from \cite{Fuku:02}, where $h=0.72$ is the Hubble parameter in $100 \; \mathrm{km}\, \mathrm{s}^{-1} \mathrm{Mpc}^{-1}$) and secondly we multiply with the ratio of B-band luminosity to stellar mass for the Milky Way (where $M_{\mathrm{MW},*}/L_{\mathrm{MW},B}=2.78M_{\odot}/L_{B,\odot}$ is taken from \cite{Portinari:04}). For the matter density distribution we used:
\begin{equation}
 \centering
\rho_{\mathrm{MW}}(R,z)=\frac{\Sigma_d}{2 z_d}\exp(-\frac{R}{R_d}-\frac{|z|}{z_d}).
\end{equation}

Putting these parts together, Eq.~(\ref{eq:rsn}) now provides us with an expression that describes the physical distribution of supernovae in the Milky Way through a mass density function, so that one could in principle combine $n_{\beta}$ with a detailed description of the neutrino distribution.

Specifically our choice of density function is model 8 of Dehnen {\it et al.}, \cite{Dehnen:97}, and we have used only the thin disk distribution (which contains most of the matter of the galaxy). The following numbers describe the the scale length, $R_d$, scale height, $z_d$, and central surface density, $\Sigma_d$:
\begin{equation}
\centering
R_d=2400\mathrm{pc}, \;\;\; z_d=180\mathrm{pc},\; \;\; \Sigma_d=1127.5 \mathrm{M}_{\odot}\mathrm{pc}^{-2}
\end{equation}
In total, the star formation and supernova rates are given by:
\begin{eqnarray}
 \centering
\psi_{\mathrm{MW}}(R,z,t)& = & 1.70\times 10^{-11} \rho_{\mathrm{MW}}(R,z)\exp[(t_0-t)/\tau] \; \\
R_{\mathrm{SN}}(R,z,t)& = &7.26\times 10^{-14} \rho_{\mathrm{MW}}(R,z)\exp[(t_0-t)/\tau].
\end{eqnarray}
Turning now to the second ingredient of Eq.~(\ref{eq:abund}) -- the amount of target material produced per supernova, $N_{\mathrm{target, SN}}$ -- we have used the nucleosynthesis simulations of Rauscher {\it et. al.} \cite{Rauscher:02} to calculate the average number of target atoms (as well as daughter atoms) produced by supernovae in the mass range 10 to 25 $M_{\odot}$:
\begin{eqnarray}
\centering
N_{\mathrm{Re,SN}}&=&2.35\pm0.99\cdot10^{22}N_A \nonumber\\
N_{\mathrm{Os,SN}}&=&5.60\pm2.35\cdot10^{21}N_A \nonumber
\end{eqnarray}

\begin{figure}
\centering
\includegraphics[scale=0.4]{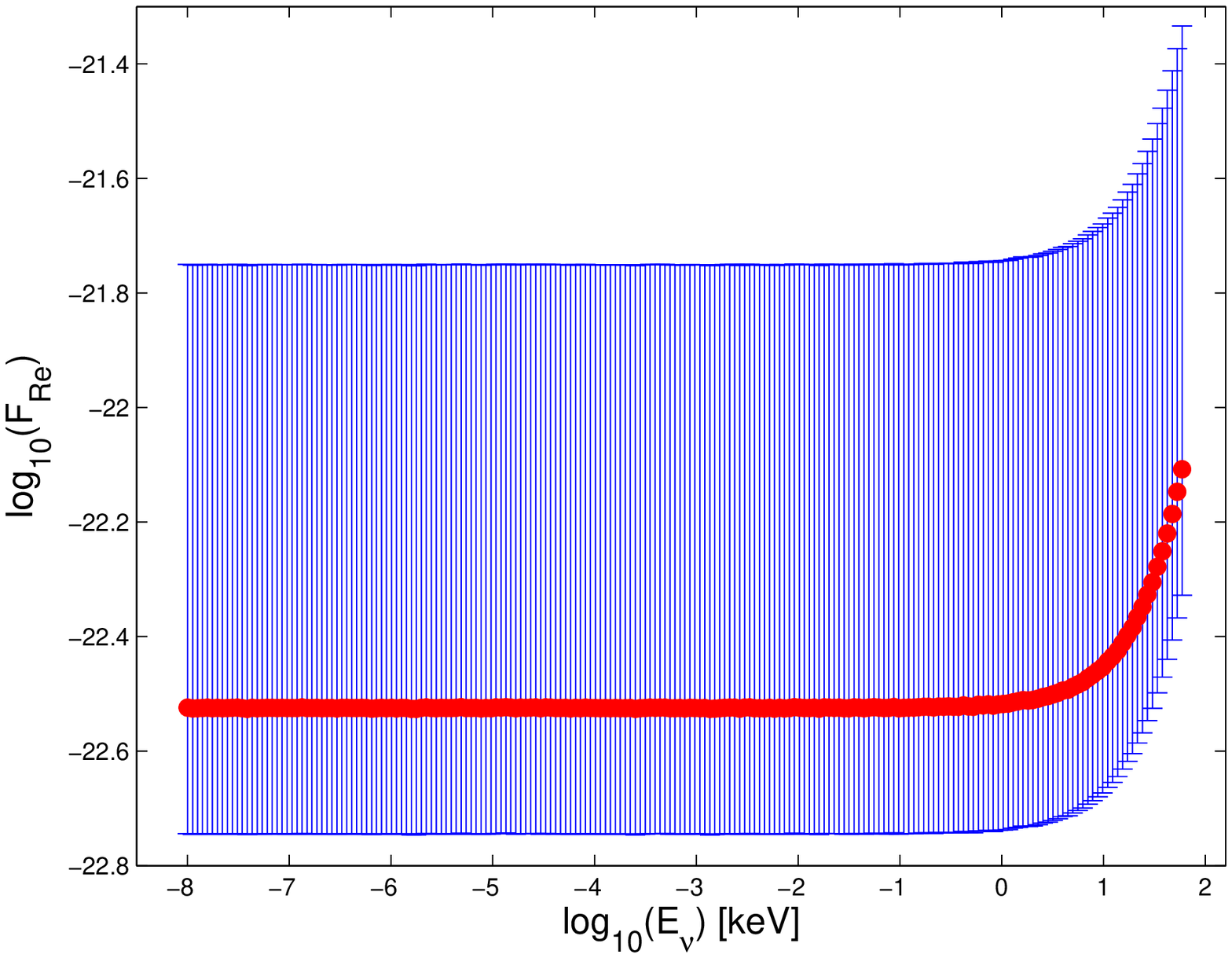}
\hfill
\includegraphics[scale=0.4]{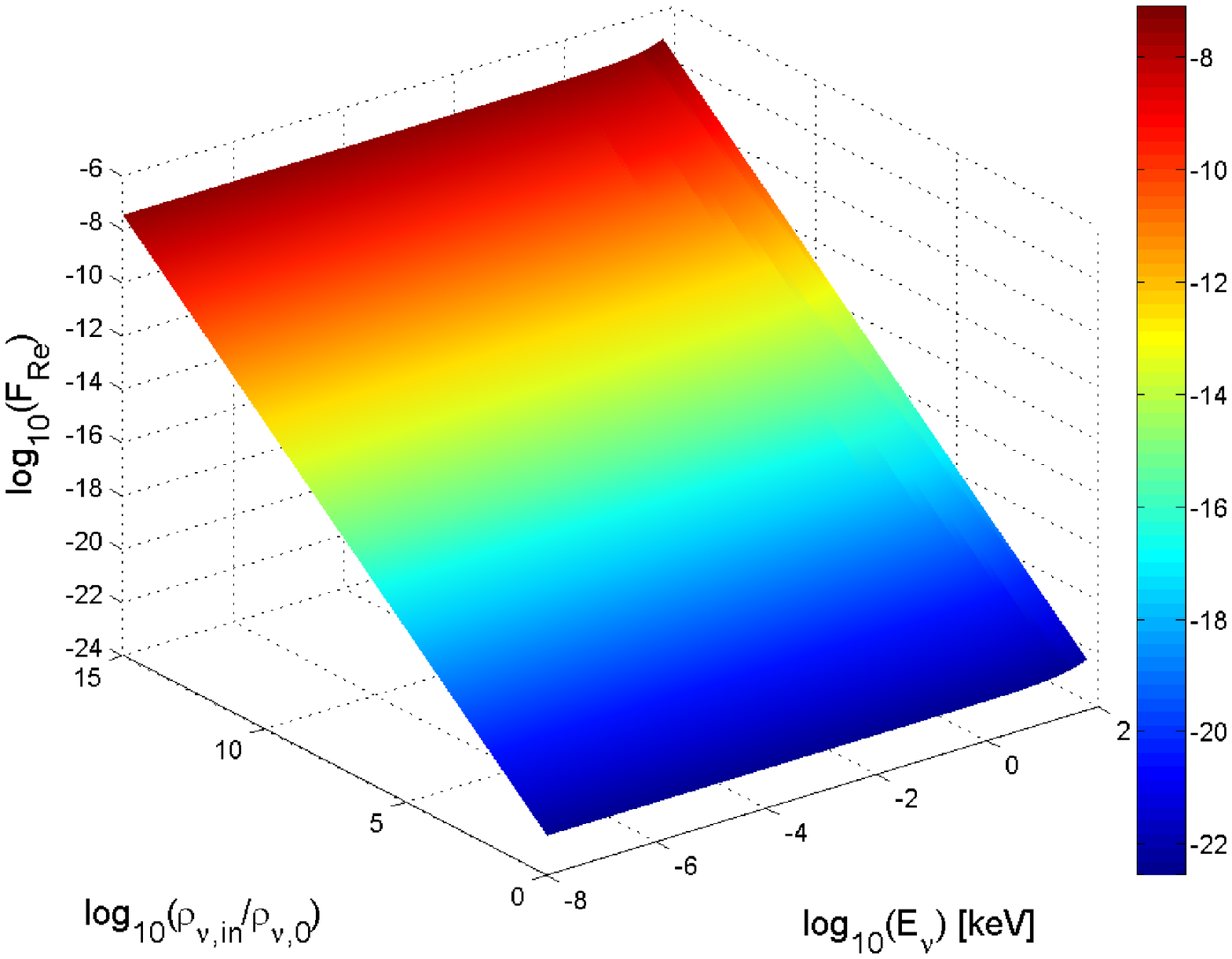}
\caption{The Rhenium abundance enhancement as a consequence of the relic neutrino background. The left panel shows our standard result, while the right panel demonstrate the expected linear enhancement of $F_{\mathrm{Re}}$ as a consequence of a larger neutrino background -- in the graph $\rho_{\nu,in}$ indicates the input neutrino density, while $\rho_{\nu,0}$ is the theoretically expected density corresponding to the normal $n_{\nu + \overline{\nu}}$=112 cm$^{-3}$. Due to the competition with coherent scattering processes, we can only assume the semi-constant value of $F_{Re}$ at low neutrino energies to be relevant. Clearly, the standard result of $F_{\mathrm{Re}}\approx10^{-22}$ is not within reach of current abundance measurements and one would need a neutrino background enhancement of at least $\mathcal{O}(10^{15})$ in order to actually see the effect of relic neutrinos on the Osmium to Rhenium abundance. It can, however, be noted that the error-bars of the plot on the left allows for an improvement of approximately one order of magnitude over our mean result. \label{fig:abbus}}
\end{figure}

Finally we have calculated the unperturbed abundances which is simply the time and volume integrated product of $n_{\beta,\mathrm{daughter}}$:
\begin{equation}
 \centering
A_{X, tot}= \int N_{X,SN} R_{SN} \mathrm{d}z \mathrm{d}R \mathrm{d}t,
\end{equation}
Taking $n_{\nu}= \frac{1}{2}n_{(\nu + \overline{\nu}),\mathrm{Universe}}$ to be a completely uniform distribution is not quite correct. But as can be seen in e.g.~\cite{Ringwald:04} the distribution for a given neutrino mass -- or analogously momentum -- is roughly constant in the inner part of the galaxy, which is also where one expects most of the matter to be. With these conventions it is very easy to calculate the abundance ratio enhancement in the presence of a neutrino density enhancement:
\begin{eqnarray}
\centering
F_{\mathrm{Re}}&=&\frac{A_{\mathrm{Os_{new}}}}{A_{\mathrm{Re_{new}}}}-\frac{A_{\mathrm{Os_{old}}}}{A_{\mathrm{Re_{old}}}}=\frac{A_{\mathrm{Os}}+ A_{\mathrm{Re+}}}{A_{\mathrm{Re}}-A_{\mathrm{Re+}}} - \frac{A_{\mathrm{Os}}}{A_{\mathrm{Re}}}, \nonumber \\
\end{eqnarray}
We present our final results in Figure~\ref{fig:abbus}.

\section{Conclusion}
As shown in Figure~\ref{fig:abbus} the abundance enhancement is tiny and far below the precision of modern galactic abundance measurements even when using meteoritic samples. Our results also show that one must have a background enhancement of around $10^{15}$ to see the effect even on the 8th decimal which is just around the precision available on the Osmium abundance -- see e.g.~\cite{Ackena:11, Lodders:03}

So despite our use of an extremely large detector volume, we must unfortunately conclude that a detection is even more orders of magnitude away with this method, than say, with a $\beta$-decay experiment such as KATRIN \cite{KAT:04} or MARE \cite{Monfardi:2006,Nucci:2010}.

\section{Acknowledgements}

We acknowledge the help in developing this idea from Hans Fynbo and useful discussions of the nuclear physics details with Karsten Riisager.

\end{document}